\documentclass{article}
\addtocounter{equation}{0}
\begin{document}
\title{Gravitational field of twisted Baby Skyrmion strings and loops}         
\author{E. \v{S}im\'{a}nek \footnote {Electronic address: simanek@ucr.edu}\\Department of Physics, University of California, Riverside, California 92521}      
\date{}          
\maketitle

\begin{abstract}

We consider the gravitational field of infinite straight and stationary twisted Baby Skyrmion cosmic string.  Using the approximate solution of Einstein equations, it is shown that the internal phase rotation (twist) along the string axis is responsible for a long-range gravitational acceleration resembling that of massive cylindrical shell.  We also study the stability and gravitational field of circular loops.  When the loop radius becomes comparable with the string width, the rigidity energy tends to stabilize small loops against radial collapse.  The nucleon scale-toroidal knot with Hopf charge $Q=1$ is found to decay very rapidly on the scale of the age of the universe due to low energy cost to flux lines crossings.  Such knot is therefore excluded from the dark matter scenario of Spergel and Steinhardt.  However, the $Q = 0$ loop, stabilized by rigidity, could be a candidate for this scenario.  In contrast, the electroweak strings are prevented from intercommuting due to much larger energy cost to intersection.  This makes them a possible candidate for the solid dark matter scenario of Bucher and Spergel.

\end{abstract}

PACS number(s):  11.27.+d, 12.39.Dc, 98.80.Cq, 3.75.Lm
\\[10pt]

\section{Introduction}

	A closed twisted Baby Skyrmion is a topological soliton in a $O(3)$ sigma model including the fourth order derivative Skyrme term.  That a stable soliton of this kind may exist was suggested some time ago by Faddeev [1].  This model is widely known as a Skyrme-Faddeev model.  More recently, it has been proposed that this model arises as a dual description of strongly coupled $SU(2)$ Yang-Mills theory [2], in which the knotted strings may represent glueballs.  The gravitational field of such glueballs may be of interest in view of the possibility that they can be a source of dark matter [3].  This is also suggested by previous investigations of cosmological implications of vortons [4].  These are loops of superconducting cosmic strings [5] stabilized against collapse by angular momentum of its charge carriers.  We note that cosmic vortons in electroweak scale were suggested as a source cold dark matter in Ref. [6].

	In the present paper we focus on the long-range gravitational force which a twisted Baby Skyrmion string exerts on surrounding matter.  We first calculate the active gravitational mass of a straight string (Secs. 1-6). The circular loop formed by bending the straight string is studied in Secs. 7 and 8.  The gravitational field of the loops is studied in Sec. 9 and Sec. 10 discusses possible role of these loops in the cold dark matter problem.

	Recently [7], we have studied the effect of internal time-dependent phase rotation on the gravitational properties of a nonlinear sigma-model with the single-axis anisotropic potential.  The precession induced by this potential prevents the lump from collapsing to a point.  This method of stabilizing the lump has been proposed by Leese [8] in a work yielding analytical solution for the fields in the $CP^1$ formulation.  Evaluating the energy-momentum tensor with the use of this solution, and solving the Einstein equations in the weak field approximation, we find that the time-dependent internal space rotation is responsible for a long-range gravitational acceleration similar to that of a rotating cylindrical shell.  Verbin and Larsen [9] studied previously this problem numerically beyond the weak coupling approximation and found that, owing to the finite frequency of spinning, the metric $g_{00}$ is not asymptotically flat consistent with the long-range gravitational potential found analytically in our paper [7]. For a string, situated parallel to the $x^{3}$-axis, this result is expected by recalling the breakdown of Lorentz-boost invariance in the $x^{3}-$ direction caused by the time-dependence of spinning-string fields [7,10]. 

Obviously, this invariance can be also lost if the string fields acquire, while remaining time-independent, an $x^{3}-$dependent phase factor.  The resulting metric $g_{00}$ is then expected to yield a long-range gravitational potential.  The numerical results of Verbin and Larsen [9] are consistent with this expectation.

Our string model is obtained by trivially extending the 2+1 dimensional Baby Skyrme model [11] to 3+1 dimensions.  The static version of this string model has been thoroughly studied in connection with the two-component Bose-Einstein condensates by Cho et al [12].  In the time-independent version of the two-dimensional Baby Skyrmion model, the size instability of the lump is prevented by the presence in the Lagrangian of the Skyrme term which is fourth order in the field variables.  For the role, this term plays in Derrick's scaling argument for stability, we refer to the book by Manton and Sutcliffe [13].  

\section{Lagrangian and Stress-Energy Tensor}

We write the Lagrangian density in the form [12] $\hbar=c=1$

\begin{equation}\label{Eq1}
\pounds = \eta^{2} \Bigr[\frac {1}{4} g ^{\mu \nu} \partial_{\mu} \vec{\phi}. \partial_{\nu} \vec{\phi} - \frac{\kappa^{2}}{8} (\partial_{\mu} \vec{\phi} \times \partial_{\nu} \vec{\phi}). (\partial ^{\mu} \vec{\phi} \times \partial ^{\nu} \vec{\phi})\Bigr]
\end{equation}

where $\mu, \nu = 0,1,2,3$.  The basic field is a triplet $\vec{\phi} \equiv (\phi_{1}, \phi_{2}, \phi_{3})$ of real scalar fields satisfying the constraint $\vec{\phi} . \vec{\phi}= 1$.  The first term in Eq. (1) is the Lagrangian density of the pure sigma-model string.  The second term is the planar analog of the three dimensional Skyrme term.  We note that, in contrast with the original Baby Skyrmion model [11,13], the anisotropic mass term is not included in Eq. (1).  As shown below, in the present model a mass term is generated by the twist of the field.  For a straight string along the $x^{3}-$ axis, we make the following Ansatz [12] for the field

\begin{equation}\label{Eq2}
\vec{\phi}= \pmatrix {\sin f (r) \cos (n \varphi + m k x^{3}) & \cr \sin f (r) \sin (n \varphi + m k x ^{3}) & \cr \cos f (r)}
\end{equation}

where $m$ and $n$ are integers, $2 \pi /k$ is the period in the $x^{3}$-direction and $r = [(x^{1})^{2} + (x^{2})^{2}]^{\frac{1}{2}}$.

The stress-energy tensor is obtained from the formula [7]

\begin{equation}\label{Eq3}
T_{\mu\nu} = \frac{2}{\sqrt {|g|}} \frac{\partial}{\partial g ^{\mu\nu}}(\sqrt {|g|\pounds})= 2 \frac{\partial \pounds}{\partial g^{\mu\nu}} - g_{\mu\nu} \pounds
\end{equation}

Using the Lagrangian density (1), the (right hand side) rhs of Eq. (3) yields

\begin{equation}\label{4}
T_{\mu\nu} = \frac{1}{2}\eta^{2}\bigl{[}\partial_{\mu}\vec{\phi}. \partial_{\nu}\vec{\phi} - \kappa^{2}(\partial_{\mu}\vec{\phi} \times \partial_{\beta}\vec{\phi}) . (\partial_{\nu}\vec{\phi} \times \partial_{\delta}\vec{\phi}) g^{\beta\delta}\bigr{]} - g_{\mu\nu} \pounds 
\end{equation}

The metric respecting the cylindrical symmetry of the string, and the twist along the $x^{3}$-axis is written as

\begin{eqnarray}\label{5}
ds^{2} = g_{00} (r) (dx^{0})^{2} - \Omega^{2} (r) \Big[(dx^{1})^{2} + (dx^{2})^{2}\Big]\nonumber \\* + 2 \Big [ g_{31}(\vec{x})dx^{3} dx^{1} + g_{32} (\vec{x}) dx^{3}dx^{2}) \Big ] + g_{33} (r)(dx^{3})^{2}
\end{eqnarray}

The components of the metric tensor are obtained by solving the Einstein equation [14]

\begin{equation}\label{Eq6}
R_{\mu\nu} = - 8 \pi G \Big(T_{\mu\nu} - \frac{1}{2} g _{\mu\nu} T \Big)
\end{equation}

where $R _{\mu\nu}$ is the Ricci tensor, $T_{\mu\nu}$ is the stress-energy tensor and $T = T ^{\mu}_{\mu}$.

In the weak field approximation, the stress-energy tensor is obtained from Eq. (4) by using the flat space metric $\eta_{\mu\nu} =$ diag $(1, -1, -1,-1)$.  Defining the quantity

\begin{equation}\label{Eq7}
S_{ij} = (\partial_{i} \vec{\phi} \times \partial_{j} \vec{\phi})^{2} 
\end{equation}

we have

\begin{equation}\label{Eq8}
T_{00} = - \pounds  
\end{equation}

\begin{equation}\label{Eq9}
T_{11} = \frac{1}{2} \eta^{2}[ (\partial_{1} \vec{\phi})^{2} + \kappa^{2} (S_{12} + S_{13})] + \pounds 
\end{equation}

\begin{equation}\label{Eq10}
T_{22} = \frac{1}{2} \eta^{2}[ (\partial_{2} \vec{\phi})^{2} + \kappa^{2} (S_{21} + S_{23})] + \pounds 
\end{equation}

\begin{equation}\label{Eq11}
T_{33} = \frac{1}{2} \eta^{2}[ (\partial_{3} \vec{\phi})^{2} + \kappa^{2} (S_{31} + S_{32})] + \pounds 
\end{equation}

\section{Active Gravitational Mass}

We first consider the 00-component of Eq. (6).  We write $g_{\mu\nu} = \eta_{\mu\nu} + h_{\mu\nu}$.

In the weak field limit, $h_{\mu\nu} << g_{\mu\nu}$, we have

\begin{equation}\label{Eq12}
R_{00}= - \frac{1}{2}(\partial ^{2}_{1} + \partial ^{2}_{2}) h_{00} = - \frac{1}{2} \bigtriangledown^{2}h_{00}
\end{equation}

On the rhs of Eq. (6), we need an expression for $\tau = T_{00} - \frac{1}{2}T$. Using Eqs. (1) and (8-11), we obtain

\begin{equation}\label{Eq13}
\tau = \frac{ \eta^{2} \kappa^{2}} {4}\Big (  S_{12} + S _{13} + S_{23}\Big)
\end{equation} 

With the Ansatz (2), we obtain from Eqs. (7) and (13)

\begin{equation}\label{Eq14}
\tau (r) = \frac{1}{4} \eta^{2} \kappa^{2}(\frac{n^{2}}{r^{2}} + m^{2}k^{2})(f' \sin f)^{2}
\end{equation}

where $f' = df/dr$. We note that this result is correct only in the presence of the twist.  If $mk = 0$, the model described by Eq. (1) is unstable unless it is augmented by an anisotropic mass term (see Ref. [13]).

Using Eqs. (6), (8) and (14), the Einstein equation for $h_{00}$ becomes

\begin{equation}\label{Eq15}
\bigtriangledown^{2} h_{00}= 16 \pi G \tau (r) 
\end{equation}

This equation can be cast in the form of a Newton-Poisson equation for the gravitational acceleration $\vec{g}= - \vec {\bigtriangledown} \phi_{g}$, where the gravitational potential $\phi_{g}$ is given $\phi_{g} = \frac{1}{2} h_{00}$ [7].  Thus, we get from Eq. (15)

\begin{equation}\label{Eq16}
\vec{\bigtriangledown} . \vec{g} (r)= -4 \pi G \rho_{a} (r) 
\end{equation}

where $\rho_{a}(r)= 2 \tau (r)$ is the active gravitational mass density.  Equation (16) can be solved for the radial component $g (r)$ of $\vec{g} (r)$ using the Gauss' law yielding

\begin{equation}\label{Eq17}
g(r) = - \frac{8\pi G}{r} \int^{r}_{0} \tau (r) r dr
\end{equation}

The profile function $f(r)$ satisfies a nonlinear differential equation which is obtained by minimizing the static energy 

\begin{equation}\label{Eq18}
\epsilon = 2 \pi \int^{\infty}_{0} T_{00} (r)r dr
\end{equation}

Using Eqs. (1), (2) and (8), we have

\begin{equation}\label{Eq19}
T_{00}(r) = \frac{1}{4} \eta^{2} \Big[f'^{2} + \Big(\frac{n^{2}}{r^{2}} + m^{2} k^{2}\Big)\Big(\sin^{2} f \Big) \Big(1 + \kappa^{2} f'^{2}\Big)\Big]
\end{equation}

This leads to the following variational equation for the profile function

\begin{eqnarray}\label{Eq20}
f'' + \frac{1}{r} f' - \Big(\frac{n^{2}}{r^{2}} + m^{2} k ^{2} \Big )\Big[ \frac{1}{2} \sin 2 f - \kappa^{2}\Big (\frac{1}{2}f '^{2} \sin 2 f + f '' \sin ^{2}f \Big )\Big] \nonumber \\* - \kappa^{2} \Big(\frac{n^{2}}{r^{2}} - m^{2} k^{2}\Big) \frac{ f' \sin^{2} f}{r} = 0
\end{eqnarray}

In what follows, we are interested in the asymptotic behaviour of the function (17) for large $r$.  If the function $\tau (r)$ exhibits an exponential decay, then for $r >> l_{D}$ where $l_{D}$ is the decay length, the upper limit of the integral on the rhs of Eq. (17) can be replaced by infinity.  In this case, the acceleration $g (r)\sim 1/r$ which corresponds to the long-range gravitational field of a massive rod.  Owing to the boundary condition $f (\infty) = 0$, the values of $f (r)$ for large $r$ are small and satisfy the linearized form of Eq. (20) given by

\begin{equation}\label{Eq21}
r^{2} f '' + r f' - (n^{2} + m^{2}k ^{2} r^{2}) f= 0 
\end{equation}

The solution of this equation is the modified Bessel function of order $n$

\begin{equation}\label{Eq22}
f (r) \sim K_{n} (m k r)
\end{equation}

The leading term of the asymptotic expansion of this function for large $r$ is 

\begin{equation}\label{Eq23}
f (r) \sim \frac{\underline{A}}{\sqrt{r}} \exp (-mkr)
\end{equation}

Hence, the decay length $l_{D} \sim 1/mk$ and the asymptotic form of Eq. (17), valid for $r >> 1 / mk$, is

\begin{equation}\label{Eq24}
g (r) = \frac {2 G m_{a}}{r}
\end{equation}

where

\begin{equation}\label{Eq25}
m_{a} \simeq 4 \pi \int ^{\infty} _{0} \tau (r)r d r = \pi \eta^{2} \kappa^{2} \int ^{\infty}_{0}\Big ( \frac{n^{2}}{r^{2}} + m^{2}k^{2}\Big) \Big( f' \sin f \Big)^{2} rdr
\end{equation}

This expression for the active gravitational mass can be rewritten in a form showing that $m_{a}$ is proportional to $m^{2}k^{2}$.  In the next section, we establish the following condition for the size stability of the string

\begin{equation}\label{Eq26}
\kappa^{2} \int^{\infty}_{0} \Big ( \frac{n^{2}}{r^{2}} f'^{2} \sin^{2} f \Big ) rdr = m^{2}k^{2} \int^{\infty}_{0} \Big ( \sin^{2}  f\Big)r dr
\end{equation}

Using this relation in Eq. (25), we obtain

\begin{equation}\label{Eq27}
m_{a} \simeq \pi \eta^{2}m^{2} k^{2} \int^{\infty}_{0}\Big(1 + \kappa^{2}f'^{2}\Big) r \sin^{2} f dr 
\end{equation}

\section{Spatial Rescaling and Stability Condition}

Starting from Eqs. (1) and (8), we write the energy (per unit length in the $x^{3}$ direction)

\begin{eqnarray}\label{Eq28}
\epsilon = \int T_{00} d^{2}x  \nonumber \\* = \frac{\eta^{2}}{4} \int \Big[ ( \partial _{1} \vec{\phi})^{2} + ( \partial _{2} \vec{\phi})^{2} +( \partial _{3} \vec{\phi})^{2} + \kappa^{2}(S_{12}+ S _{13}+S_{23}) \Big]d^{2} x
\end{eqnarray}

We now consider the transformation of this expression under a spatial rescaling in the $x^{1}, x^{2}$-space [13]

\begin{equation}\label{Eq29}
x^{1} \rightarrow y^{1} = \mu x^{1}, x^{2}\rightarrow y ^{2}= \mu x^{2}, x ^{3}\rightarrow y^{3}= x^{3}, d^{2}x = \frac{1}{\mu^{2}} d^{2}y  
\end{equation}

This implies a following transformation for the fields

\begin{equation}\label{Eq30}
\partial_{1}\vec{\phi}(\vec{x}) \rightarrow \mu \frac{\partial \vec{\phi} (\vec{y})}{\partial y^{1}},
\partial_{2}\vec{\phi} \rightarrow \mu \frac{\partial \vec{\phi} (\vec{y})}{\partial y^{2}},
\partial_{3}\vec{\phi} \rightarrow  \frac{\partial \vec{\phi} (\vec{y})}{\partial y^{3}}
\end{equation}

For the Skyrme terms, we obtain

\begin{equation}\label{Eq.31}
S_{12} = \Big(\partial_{1} \vec{\phi} \times \partial _{2}\vec{\phi}  \Big)^{2} \rightarrow \mu^{4} \Big(\frac{\partial \vec{\phi} (\vec{y})}{\partial y^{1}}\times \frac{\partial \vec{\phi} (\vec{y})}{\partial y^{2}}  \Big )^{2}= \mu ^{4} S_{12}(\vec{y})
\end{equation}

and 

\begin{equation}\label{Eq.32}
{S_{13} \rightarrow \mu^{2} S_{13}(\vec{y})\atopwithdelims \{\} S_{23}\rightarrow \mu^{2} S_{23}(\vec{y})}\qquad
\end{equation}

Applying the maps (29)-(32) in Eq. (28), we obtain the rescaled energy

\begin{eqnarray}\label{Eq.33}
\epsilon (\mu) = \frac{\eta^2}{4} \int \Big[ \Big( \frac{\partial \vec{\phi}}{\partial y^{1}} \Big)^{2} + \Big( \frac{\partial \vec{\phi}}{\partial y^{2}} \Big)^{2} + \frac{1}{\mu^{2}} \Big( \frac{\partial \vec{\phi}}{\partial y^{3}} \Big)^{2} +\kappa^{2}\mu^{2} S _{12} (\vec{y}) \nonumber \\* + \kappa^{2} S_{13} (\vec{y}) +  \kappa^{2}S_{23} (\vec{y}) \Big]d^{2} y
\end{eqnarray}

From the condition that energy is stationary under spatial rescaling, we obtain

\begin{equation}\label{Eq34}
\frac{d \epsilon (\mu)}{d \mu} = - \frac{2}{\mu^{3}} \int ( \partial_{3} \vec{\phi})^{2} d ^{2} y + 2 \mu \kappa^{2} \int S_{12}(\vec{y}) d ^{2}y = 0
\end{equation}

which implies

\begin{eqnarray}\label{Eq35}
\int \Big(\partial_{3} \vec{\phi} \Big)^{2} d^{2}y = \kappa^{2} \int \Big ( \partial_{1} \vec{\phi} \times \partial_{2} \vec{\phi}\Big )^{2}d^{2} y
\end{eqnarray}

Using Ansatz (2) for the fields, the stability condition (35) takes the form given in Eq. (26).

\section {Relation of String Width to Twist Rate}

A qualitative understanding of the profile funcion $f (r)$ can be obtained by considering Eq. (20) for $r \rightarrow 0$.  Introducing the Ansatz

\begin{equation}\label{Eq36}
f (r) = \pi - Cr^{p}  
\end{equation}

into Eq. (20) and keeping only the leading terms for $r \simeq 0$, we have

\begin{equation}\label{Eq37}
C \Big ( n^{2}- p^{2}\Big ) r ^{p-2} + C m^{2} k^{2}r ^{p} - 2 \kappa ^{2}C ^{3}n ^{2} \Big(p^{2}- p \Big) r ^{3 p-4} = 0
\end{equation}

For $n = 1$, we obtain $p = n =1$.  In what follows, we confine ourselves to the case $n =2$, since it allows us to make further progress without resorting to numerical work.  In this case, the first term in Eq. (37) implies $p =n = 2$.  The vanishing of the sum of the second and third terms yields for $p =2$

\begin{equation}\label{Eq38}
C^{2} = \frac{m^{2}k^{2}} {16 \kappa^{2}}
\end{equation}

Hence, the profile function for $n =2$ goes as

\begin{equation}\label{Eq39}
f (r) = \pi - \frac{mk}{4 \kappa} r^{2}
\end{equation}

This suggests that the width $\lambda$ is related to the twist rate $m k$ as follows

\begin{equation}\label{Eq40}
\lambda^{2} \sim \frac{\kappa}{m k} 
\end{equation}

To substantiate this prediction, we make a variational Ansatz for $f(r)$ of the form

\begin{equation}\label{Eq41}
f (r) = \cos ^{-1} \frac{r^{4} - \lambda ^{4}}{r^{4}+ \lambda^{4}}
\end{equation}

where $\lambda$ is the variational parameter to be determined from the stability condition (26).  We note that Eq. (41) describes the $n = 2$ lump solution of radius $\lambda$ in the sigma model.  Expanding the function (41) into Taylor series about $r = 0$, we have

\begin{equation}\label{Eq42}
f (r)\simeq \pi - \frac{2 r ^{2}}{\lambda ^{2}}
\end{equation}

Comparing this result with the Eq. (39), we obtain $\lambda^{2} = 8 \kappa/m k$.

We now turn to the variational estimate of $\lambda$.  According to the stability condition (26), the soliton has a preferred size at which the contribution of the Skyrme term is equal to the effective anisotropy energy induced by the twist.  Introducing the Ansatz (41) into Eq. (26), and performing the $r-$ integrations, we obtain the following results

\begin{equation}\label{Eq43}
\kappa^{2} n^{2} \int^{\infty}_{0} \frac{f^{'2} \sin^{2}f}{r^{2}} rdr = \frac {\pi \kappa^{2} n^{2}}{ \lambda ^{2}}
\end{equation}

and

\begin{equation}\label{Eq44}
m ^{2} k^{2} \int^{\infty}_{0}r \sin^{2} f dr = \frac {m^{2}k ^{2} \pi\lambda^{2}}{2}
\end{equation}

We see that the Skyrme contribution (43) goes as $\lambda^{-2}$.  Thus it favors large values of $\lambda$.  On the other hand, the anisotropy term (44) favors small $\lambda$.  Eq. (26) describes the balance between these tendencies yielding a relation (valid for $n =2$)

\begin{equation}\label{Eq45}
\lambda^{2} = \frac {\sqrt{8 }\kappa}{m k}
\end{equation}

For $n = 1$, the profile function is

\begin{equation}\label{Eq46}
f (r) = \cos^{-1} \frac{r^{2} - \lambda^{2} }{r ^{2} + \lambda^{2}}
\end{equation}

For this function, the effective anisotropy energy diverges owing to the power law decay for large $\lambda$.  

Let us estimate the expression (27) for $n= 2$.  Using the profile function (41), we obtain

\begin{equation}\label{Eq47}
m_{a} = \pi \eta^{2} m^{2} k^{2} \Big ( \frac{\pi \lambda^{2}}{2} + \frac{8 \kappa^{2}}{3} \Big) 
\end{equation}

Now, $\lambda$ is not a free parameter, rather it is fixed to the preferred value given by Eq. (45).  Using this value in Eq. (47), the final expression for the active gravitational mass per unit strength of the string becomes

\begin{equation}\label{Eq48}
 m_{a} = \pi \eta^{2} \Big ( \pi \sqrt{2} |u| + \frac{8}{3} u^{2}\Big)
\end{equation}

where we introduced the dimensionless parameter

\begin{equation}\label{Eq49}
u= \kappa mk 
\end{equation}

\section {Limit of Pressureless String}

From the expressions (48) and (49) it is apparent that a nonzero value of $m_{a}$ is obtained only if the twist rate $m k$ and the magnitude of the Skyrme term $\kappa$ are both nonzero.  We now show that the magnitude of string pressure decreases on increasing the parameter $u$.  For a certain critical value of $u$, the pressure vanishes and the string turns into a pressureless gas in analogy with the vortons [15].

For a string situated parallel to the $x^{3}$-axis, the pressure in the $x^{3}$-direction is given by

\begin{equation}\label{Eq50}
p = \int T_{33} d^{2}x 
\end{equation}

Using Eqs. (1) and (11), we have

\begin{equation}\label{Eq51}
T_{33} = \frac{\eta^{2}}{4} \Big[ - (\partial_{1} \vec{\phi})^{2} - (\partial _{2}\vec{\phi})^{2} + \kappa^{2} (S_{13} + S_{23}) - \kappa^{2} S_{12} + (\partial_{3}\vec{\phi})^{2}\Big] 
\end{equation}

Performing the transverse integral of this equation, and using the size stability condition (35), we obtain

\begin{equation}\label{Eq52}
p = \frac{-\eta^{2}}{4} \int \Big [(\partial_{1}\vec{\phi})^{2} + (\partial_{2}\vec{\phi})^{2} - \kappa^{2} (S_{13} + S_{23}) \Big]d^2x 
\end{equation}

Since the expression $\kappa^{2} (S_{13} + S_{23})$ is proportional to $u^{2}$, we see from Eq. (52) that $|p|$ decreases as $u$ increases.

Using the Ansatz (2), we have

\begin{equation}\label{Eq53}
\int \Big [ \Big(\partial_{1}\vec{\phi}\Big)^{2} +  \Big(\partial_{2} \vec{\phi} \Big)^{2} \Big] d^{2} x = 2 \pi \int ^{\infty}_{0} \Big( f'^{2} + \frac{n^{2}}{r^{2}} \sin ^{2} f \Big)r dr
\end{equation}

and

\begin{equation}\label{Eq54}
\kappa^{2} \int \Big ( S_{13} + S _{23} \Big ) d^{2} x = 2 \pi u ^{2} \int ^{\infty}_{0} \Big (f'^{2} \sin ^{2}f \Big ) r dr 
\end{equation}

Introducing Eqs. (53) and (54) into (52), we obtain the condition for vanishing pressure in a general form

\begin{equation}\label{55}
 \int ^{\infty}_{0} \Big( f^{'2} + \frac{n^{2} \sin ^{2} f}{r^{2}} \Big ) r d r = u ^{2} \int ^{\infty}_{0} \Big ( f ^{'2}  \sin ^{2} f \Big )  r d r
\end{equation}

Confining ourselves to the $n = 2$ case, we use in Eq. (55) the expression (41) and obtain the critical value of $u$ for which the pressure vanishes

\begin{equation}\label{56}
u_{crit} = \sqrt{3} 
\end{equation}

Introducing this quantity into Eq. (48), we obtain the upper bound for the active gravitational mass

\begin{equation}\label{57}
m_{a} (u = u_{crit}) = 49 \eta^{2}
\end{equation}

It is interesting to compare the expression (48) and (57) with the $u$-dependence of the energy, per unit length of the string, $\epsilon$, given by Eq. (18).  Using the size-stability condition (26), we obtain from Eqs. (18) and (19)

\begin{equation}\label{58}
\epsilon = \frac{\pi \eta^{2}}{2} \int _{0} ^{\infty} \Big( f'^{2} + \frac{n^{2}}{r^{2}} \sin^{2} f + 2 m^{2}k^{2}\sin^{2} f + \kappa^{2}m^{2} k^{2} f'^{2} \sin^{2} f \Big) r dr
\end{equation}

Performing the transverse integrations with the $n = 2$ profile function $f (r)$ (given in Eq. (41)), we obtain

\begin{equation}\label{59}
\epsilon = \eta^{2} \Big (4 \pi + \sqrt{2}\pi^{2} u + \frac{4 \pi}{3} u^{2} \Big )
\end{equation} 

For $u= u_{crit} = \sqrt{3}$, this equation yields the critical energy $\epsilon_{crit}$ corresponding to the case of vanishing string pressure.  It is interesting that for this value of $u$, the critical active gravitational mass $m_{a}$ is, according to Eq. (57), exactly equal to $\epsilon_{crit}$.  This conclusion follows more directly by considering the sum 

\begin{equation}\label{60}
\epsilon + p = \frac{\eta^2 \kappa^2}{2} \int (S_{12} + S_{13}+ S_{23})d^{2}x = m_{a}
\end{equation} 

where the first equality is the sum of expressions (28) and (52), and the second is the consequence of Eqs. (13) and (25).

\section {Stability and Radius of a Circular Loop}

Consider a circular loop of radius $R$ much larger than the thickness of the string $\lambda$.  The loop is created by bending the straight string and connecting smoothly the periodic ends.  A large loop is stabilized by the Skyrme energy and the twist along the $x ^3$-axis.  This leads to a repulsive centrifugal stress opposing the centripetal force due to the tension. For small loops $(R/\lambda \simeq 1)$, the ridigity due to elastic bending energy provides an additional loop stabilization mechanism.  Discussion of this effect is postponed to Sec. 8.

To study the classical stability of the loop, we confine ourselves to the $n=2$ case and use Eq. (59) to write the energy of the loop in the form

\begin{equation}\label{61}
\epsilon_{loop} = 2 \pi R \epsilon = 2 \pi R \eta^{2} \Big [ 4 \pi + \pi ^{2} \sqrt{2} m k \kappa + \frac{4 \pi}{3} \Big ( m k \kappa \Big )^{2} \Big ]
\end{equation} 

where the definition $u = m k \kappa$ is invoked.  To proceed, we need to relate the linear momentum $mk$ to the radius $R$ of the loop.  From Eq. (2), we see that the transverse fields pick up a phase factor $mk$ per unit length along the coordinate $x^{3}$.  Thus, if we denote by $\vartheta$ the phase factor for length $x^{3}$, we have $m k = \partial_{3} \vartheta$.  Next, we relate the winding number density $\partial_{3} \vartheta$ to the topological number $N$ defined as [5]

\begin{equation}\label{62}
N = \frac{1}{2 \pi} \oint d l \partial _{l} \vartheta
\end{equation} 

For a circular loop of radius $R$, we obtain from Eq. (62) $N = R \partial_{3} \vartheta$ yielding

\begin{equation}\label{63}
m k = \frac{N}{R}
\end{equation} 

Introducing this relation into the Eq. (61), we obtain the loop energy as a function of the loop radius

\begin{equation}\label{64}
\epsilon_{loop} = 8 \pi^{2} \eta^{2} \Big (R + \frac{\kappa^{2} N^{2}}{3 R} \Big) + K
\end{equation} 

where $K$ is the $R$-independent contribution given $K = 2^{\frac{3}{2}} \pi^{3}\eta^{2}N \kappa$.  The expression (67) has a minimum at the preferred radius $R_{e}$ given by

\begin{equation}\label{65}
R_{e}^{2} = \frac{\kappa^{2}N^{2}}{3}
\end{equation} 

Using Eqs. (63) and (65), the equilibrium value of the parameter $u$ becomes

\begin{equation}\label{66}
u_{e} = \frac{N \kappa}{R_{e}} = \sqrt {3}
\end{equation} 

Recalling Eq. (56), we see that $u_{e} = u_{crit}$.  Thus, the equilibrium value of $u$ that ensures the stability of the ring is equal to the critical value $u$ for which the pressure $p$ vanishes.   

We now consider stability of a loop of small radius $R_{e}$.  In this case, there is a problem that is revealed by considering the ratio $R_e / \lambda$.  Using Eqs. (45) and (65), we obtain $R_e / \lambda \simeq 0.45 N$.  This indicates that, for $N=1$, the loop tends to collapse under the centripetal force due to string tension. 

\section {Rigidity of Small Loops}

In what follows, we show that an enhancement of $R_e / \lambda$ is obtained from a model in which the string energy is augmented by a contribution due to rigidity.  To estimate this contribution, we follow the analogy with elasticity theory of bending rods [16].  Thus, we picture the rigid finite width-string as a bundle of filamentary lines each representing a flexible zero width-string.

Due to bending, lines on the convex side of the string are extended, whereas those on the concave are compressed.  These deformations are described by the relative extension $(d x ^{3'} - d x ^{3})/ dx^{3} = u_{33}$ where $u_{33}$ is the strain tensor.  For a simple extension, we can use Hooke's law to find the stress tensor[16]

\begin{equation}\label{67}
\sigma_{33}(r,\phi) = T_{00}(r)u_{33} (r,\phi)
\end{equation} 

where, $r, \phi$ are the polar coordinates in the cross section of the string.  The elastic energy per unit volume of the string is

\begin{equation}\label{68}
V (r, \phi) = \frac {1} {2} \sigma_{33} u_{33} = \frac{1}{2} T_{00} (r) u^{2}_{33} (r, \phi)
\end{equation} 

Integrating this expression over the cross section, we obtain the bending energy per unit length of the string

\begin{equation}\label{69}
\epsilon_{b} = \frac{1}{2} \int^{\infty}_{0} rdr \int ^{2 \pi}_{0} d \phi T _{00}(r)u^{2}_{33} (r, \phi)
\end{equation} 

Evaluating the angular integral of $u^{2}_{33}$ with use of the complete elliptic integral of second kind and limiting ourselves to the lowest power in $r/R$, we obtain

\begin{equation}\label{70}
I (r) = \int ^{2\pi}_{0} d \phi u^{2} _{33} (r, \phi) \simeq \frac{\pi r^{2}}{R^{2}}
\end{equation} 

Introducing this result into Eq. (69), the bending energy becomes

\begin{equation}\label{71}
\epsilon_{b} \simeq \frac{1}{2} \int r I (r) T _{00} (r) dr = \frac{\pi}{2 R ^{2}} \int ^{\infty}_{0} T _{00} (r) r^{3} dr
\end{equation} 

The bending energy leads to a modification of the transverse stability condition (35).  As a consequence, the equation (45) is also modified leading to a smaller value for the string  width.  Applying the spatial rescaling (29) to the energy $\epsilon + \epsilon _{b}$, we obtain

\begin{eqnarray}\label{72}
\int^{\infty}_{0} (\partial_{3} \vec \phi)^{2} rdr = \kappa ^{2} \int^{\infty}_{0} (\partial_{1} \vec \phi \times \partial_{2} \vec \phi)^{2} rdr \nonumber \\* - \frac{1}{4R^{2}}  \int^{\infty}_{0} \Big[ (\partial_{1} \vec \phi)^{2}+ (\partial_{2}\vec \phi)^{2} + \kappa^{2} (S_{13}+ S_{23}) \Big] r^{3}dr
\end{eqnarray} 

where the second integral on the rhs represents the modification of the condition (35) caused by bending. We have neglected the contribution to this integral due to the term $(\partial_{3}\vec{\phi})^{2}$.  This contribution is logarithmically divergent and a cutoff radius of order $R$ must be imposed.  Then it turns out that its magnitude is less than 1/10 of that due to the second integral on the rhs of Eq. (72). Evaluating the integrals in Eq. (72) for the $n = 2$ Ansatz (41), and using the condition (63) for $N=1$, we obtain the string width in the presence of a bend as a function of the loop radius

\begin{equation}\label{73}
\lambda^{2}_{b} (R) = \frac{2 \kappa R^{2}} {(\frac{3}{2} R^{2} + \frac{1}{4} \kappa^{2})^{\frac{1}{2}}}
\end{equation} 

If we compare this result with Eq. (45), we see that $\lambda_{b} < \lambda$.  This is expected as the bending energy is decreased on decreasing the width of the string.

Now, we consider the effect of the bending energy on the stability of the loop.  Substituting into the rhs of Eq. (71) the quantity $T_{00}$, and using the transverse stability condition (72), we obtain the total energy per unit length

\begin{eqnarray}\label{74}
\epsilon _{tot} = \epsilon + \epsilon _{b} = \frac {\pi \eta^{2}}{2} \int^{\infty}_{0} \Big [ (\partial_{1} \vec{\phi})^{2} \nonumber \\* + (\partial_{2} \vec{\phi})^{2} + 2 (\partial_{3} \vec{\phi})^{2} + \kappa^{2} (S_{13} + S_{23}) \Big ] rdr \nonumber \\* +  \frac{\pi \eta^{2}}{8 R^{2}} \int^{\infty}_{0} \Big \lbrace 2 \Big[ (\partial_{1} \vec{\phi})^{2} \nonumber \\* + (\partial_{2} \vec{\phi})^{2} + \kappa^{2} (S_{13} + S_{23})\Big  ] + \Big [ (\partial_{3} \vec{\phi})^{2} + \kappa ^{2} S_{12} \Big ] \Big \rbrace r^{3}dr
\end{eqnarray} 

Evaluating the integrals in this equation for $n = 2$ and $N= 1$, we are led to the following result for the total energy per unit length

\begin{eqnarray}\label{75}
\epsilon_{tot} = \frac{\pi \eta^{2}}{2} \Big ( 8 +\pi m^{2}k^{2}\lambda^{2}_{b}+ \frac {8}{3} m^{2} k^{2}\kappa^{2} \Big ) \nonumber \\* + \frac{\pi \eta^{2}}{R^{2}} \Big  ( \pi \lambda^{2}_{b} + \frac{\pi}{4} m^{2}k^{2} \kappa^{2}\lambda^{2}_{b} + \frac{4} {3} \kappa^{2} \Big ) \nonumber \\* + \frac{\pi \eta^{2}}{8 R^{2}} m^{2}k^{2}\lambda^{4}_{b} \Big[ \log (1 + \frac {R^{4}}{\lambda^{4}_{b}}) - \frac {R^{4}}{\lambda^{4}_{b}} \Big ( 1 + \frac {R^{4}}{\lambda^{4}_{b}} \Big)^{-1} \Big]
\end{eqnarray} 

To simplify numerical calculations, we neglect the small last term on the rhs of this eqation and obtain the loop energy in the form 

\begin{equation}\label{76}
\epsilon_{loop} \simeq 2 \pi R \eta ^{2}\Big [ 4 \pi + \frac {3 \pi^{2}\kappa}{(\frac{3}{2} R^{2}+ \frac {1}{4}\kappa^{2})^{\frac{1}{2}}}  \Big(  1 + \frac {\kappa^{2}}{6 R^{2}} \Big ) + \frac{8 \pi \kappa^{2}} {3 R^{2}}\Big ]
\end{equation} 

The minimum of this expression occurs for the preferred radius $R_{e}$ satisfying the following equation

\begin{equation}\label{77}
4 \simeq \frac{8 \kappa^{2}}{3 R ^{2}_{e}} + \frac {3 \pi \kappa^{3}}{4(\frac{3}{2} R^{2}_{e} + \frac{1}{4} \kappa^{2})^{\frac{3}{2}}} \Big(  1 + \frac{\kappa^{2}}{6 R^{2}_{e}}\Big)
\end{equation} 

A numerical solution of this equation yields $R^{2}_{e} \simeq \kappa^{2}$.  Using this value in Eq. (73), we obtain $\lambda^{2}_{b} \simeq 1.5 \kappa^{2}$.  From these results, we obtain the ratio $R_{e}/ \lambda_{b} \simeq 0.81$ showing a substantial enhancement in comparison with $R_{e} / \lambda \simeq 0.45$ obtained from the calculation which disregards the effect of the bending energy.

Let us compare $\epsilon _{loop}$ obtained from Eq. (76) with the numerical results obtained in Ref. [17] for the minimum energy configurations characterized by Hopf charge $Q$.  According to Ref. [12], the knot quantum number $Q = mn$.  Since $2\pi/k$ is one period in the $x^{3}$ coordinate, we have $k = 1/R$ which implies $m=N$.  Thus, the loop with $N=1, n=2$ has Hopf charge $Q=2$.

The numerical result for the $Q=2$ soliton energy is $835 $ e.u. [17].  The energy unit e.u. $ = F_{\pi}/4e = \kappa \eta^{2}/4$.  By fitting the parameters $F_{\pi}$ and $e$ to the baryon masses, Adkins, Nappi, and Witten [18] find $F_{\pi} = 129$ MeV and $e = 5.45$.  Thus the energy unit used in Ref. [17] is e.u. $ = 5.9$ MeV  and the $Q=2$ soliton energy is $E =4.92$ GeV.  This should be compared with our result for the loop energy $295 \kappa \eta^{2} \simeq 7$ GeV. This is consistent with the numerical results of Gladikowski and Hellmund [19] who also found that the energy of the configuration with $N= 2, n=1$ is lower than that with $N=1, n=2$.

\section {Gravitational Field of a Circular Loop}

Starting with the Einstein equation in three spatial dimensions, we introduce the gravitational potential $\phi_{g} = h_{00}/2$ and obtain from the 00-component of Eq. (6) a Newton-Poisson equation with the solution

\begin{equation}\label{78}
\phi_{g} (\vec{x}) = - 2 G \int \frac {\tau (\vec{x}')}{| \vec{x} - \vec{x}'|} d^{3}x'
\end{equation} 

where $\tau (\vec{x}')$ is given in Eq. (13).  Developing $1/|\vec{x} - \vec{x}'|$ in a Taylor series, we obtain

\begin{equation}\label{79}
\phi_{g} (\vec {x}) = \phi_{0} (\vec {x}) + \phi_{2} (\vec {x}) + ...
\end{equation} 

where

\begin{equation}\label{80}
\phi_{0} (\vec{x}) = - \frac{2G}{x} \int \tau (\vec{x}') d^{3} x' = - \frac{ M_{a} G} {x}
\end{equation} 

where $x$ is the distance from the loop center.

The quantity $M_{a}$ is the total active gravitational mass contained in the integration volume.  Owing to the rotational symmetry of the loop, there are no odd terms in the expansion (79).  In what follows, we assume that $x \gg R$.  Then $\phi _{0}(\vec{x})$ dominates this expansion.

For a large loop of radius $R$, we have $M_{a} = 2 \pi R m_{a}$, where $m_{a}$ is the active gravitational mass density per unit length of the string discussed in Sec. 3.  Using Ansatz (41) on the rhs of Eq. (25), we obtain

\begin{equation}\label{81}
m_{a} = \pi \eta^{2} \kappa^{2} \Big ( \frac{4 \pi}{\lambda^{2}} + \frac{8 m^{2}k^{2}}{3} \Big)
\end{equation} 

Using in Eq. (81) the equilibrium values $u_{e} = \sqrt{3}, R_{e} = N \kappa / \sqrt {3}$, and the quantity $\lambda ^{2}$ given in Eq.(45) we obtain $m_{a} \simeq 49 \eta ^{2}$. The active gravitational mass of the loop is then equal to $M_{a} = 2 \pi R_{e}m_{a} \simeq 178 N \kappa \eta ^{2}$.  

For a small loop, we need to calculate $m_{a}$ from Eq. (25) with $\tau (r)$ that is modified by the bending energy.  The gravitational mass per unit length of a rigid string is given by

\begin{equation}\label{82}
\bar {m}_{a} \simeq 4 \pi \int ^{\infty}_{0} \bar{\tau} (r)rdr = 2 \pi \int ^{\infty}_{0} (\bar{T} _{00} + \bar{T} _{11} + \bar{T} _{22} + \bar{T} _{33})rdr
\end{equation}

where $\bar{T}_{ii}$ equals $T_{ii}$ renormalized by the bending energy $\epsilon_{b}$.  Thus, we have

\begin{equation}\label{83}
2 \pi \int ^{\infty}_{0} \bar{T}_{00} rdr = 2 \pi \int^{\infty}_{0} T_{00} rdr + \epsilon _{b} 
\end{equation}

and

\begin{equation}\label{84}
2 \pi \int ^{\infty}_{0} \bar{T}_{33} rdr = 2\pi \int ^{\infty}_{0} T_{33}rdr - \epsilon_{b}
\end{equation}

Note that the lhs of Eq.(84) is the net pressure integrated over the cross section of the rigid string.  The second term on the rhs is the additional pressure due to the stress tensor $\sigma _{33}$ given in Eq.(67).  This leads to the identity

\begin{equation}\label{85}
\int _{0}^{\infty} rdr \int ^{2 \pi} _{0} \sigma _{33} (r, \phi) d \phi \simeq \frac{\pi}{2 R^{2}} \int ^{\infty}_{0} T _{00}(r) r ^{3} dr = \epsilon _{b}
\end{equation}

Next, we consider the remaining terms of Eq. (82).  Using Eqs. (9) and (10), we have

\begin{equation}\label{86}
2 \pi \int ^{\infty}_{0} \Big (\bar{T} _{11} + \bar{T} _{22} \Big ) rdr = \pi \eta^{2} \int ^{\infty} _{0} \Big [\kappa ^{2} S _{12} - (\partial _{3} \vec{\phi})^{2} \Big] rdr
\end{equation}

For a flexible string, the rhs of Eq.(86) vanishes owing to the stability condition (35).  For a rigid string, this condition is replaced by Eq.(72) so that the rhs of Eq. (86) is nonzero and given by

\begin{eqnarray}\label{87}
2 \pi\int^{\infty}_{0} \big (  \bar {T} _{11} + \bar{T} _{22}\big)rdr = \frac {\pi \eta^{2}} {4 R ^{2}} \int ^{\infty} _{0} \Big [ \big (\partial_{1} \vec{\phi} \big)^{2} \nonumber \\* + \big (\partial_{2} \vec{\phi} \big )^{2} + \kappa ^{2} \big(  S _{13} + S _{23}\big)\Big] r^{3}dr
\end{eqnarray}

Using Eqs. (11), (83), and (84), we have

\begin{equation}\label{88}
2 \pi \int^{\infty}_{0} \Big ( \bar {T}_{00} + \bar{T}_{33} \Big )r d r = \pi \eta^{2} \int^{\infty}_{0} \Big [ \big( \partial_{3} \vec{\phi} \big )^{2} + \kappa^{2} \big ( S_{13}+ S_{23} \big ) \Big] r d r 
\end{equation}

Introducing Eqs. (87) and (88) into (82), yields

\begin{eqnarray}\label{89}
\bar{m}_{a} = \pi \eta^{2} \int^{\infty}_{0} \Big [ \big (  \partial _ {3} \vec{\phi} \big )^{2} + \kappa^{2} \big ( S_{13} + S _{23} \big ) \Big ] rdr \nonumber \\ * + \frac {\pi \eta ^{2}} {4 R^{2}} \int ^{\infty} _{0} \Big [ \big ( \partial_{1} \vec{\phi} \big )^{2} + \big ( \partial_{2} \vec{\phi} \big )^{2} + \kappa ^{2} \big ( S _{13}+ S _{23} \big ) \Big ] rdr
\end{eqnarray}

Performing the integrations over $r$ with use of Ans\"{a}tze (2) and (41), Eq. (89) yields

\begin{equation}\label{90}
\bar{m}_{a} = \eta^{2} \Big [ N^{2} k^{2} \Big ( \frac{\pi ^{2}\lambda ^{2}_{b}} {2} + \frac{8 \pi \kappa^{2}} {3} + \frac{\pi ^{2} \kappa^{2} \lambda^{2}_{b}} {4 R^{2}} \Big ) \Big ] + \eta^{2} \frac{\pi^{2} \lambda^{2}}{R^{2}_{e}}
\end{equation}

From the numerical solution of Eq. (77), we have $ k^{2} \lambda _{b}^{2} = \lambda ^{2}_{b} /R_{e}^{2} \cong 1.5, k^{2} \kappa^{2} \simeq 1$, and $\lambda^{2}_{b} / \kappa^{2} \simeq 1.5$.  Introducing these values into Eq. (90), we have $\bar {m}_{a} \simeq 34 \eta^{2}$.  Consequently, the active gravitational mass of the $N = 1, n = 2$ loop, $\bar {M} _{a} = 2 \pi R _{e} \bar {m}_{a}$, for the nucleon scale is given by $\bar {M}_{a} \simeq 214 \kappa \eta ^{2} \simeq 5 $ GeV.  This is to be compared with the net mass of the $N = 1, n = 2$ loop which according to Eq. (76) amounts to $\epsilon _{loop} \simeq 295 \kappa \eta^{2} \simeq 7 $ GeV.  The decrease of $\bar {M} _{a}$ relative to $\epsilon_{loop}$ is to be attributed to the rigidity of the small loops.

\section{Knots and Dark Matter}

Knowing that the twisted Baby Skyrmion string and loops produce a nonzero long range gravitational acceleration, prompts us to look into the cosmological implications of the model studied in this paper.  In what follows, we focus on the question if the smallest rings (knots) could be a candidate for the cold dark matter.

Spergel and Steinhardt [3] proposed a scenario in which the dark matter particles are self-interacting so that they have large elastic scattering cross-section but negligible annihilation with a lifetime larger than the present age of the universe.  We note that the model described by Lagrangian (1) is also self -interacting owing to the presence of the Skyrme term that is fourth order in the field variables.  As pointed out in Ref. [12], this term is responsible for the topological stability of the rings coming from linking the two color magnetic fluxes.  In the scenario of Ref. [3], the mean-free path of the particles should be larger than 1 kpc but smaller than about 1 Mpc.  This implies a range $8 \times  10 ^{- (25 - 22)}$ cm$^{2}$/GeV for the ratio $\sigma / m $ where $\sigma$ is the scattering cross section and $m$ is the mass of the dark matter particles.

We now examine this ratio for the nucleon scale knot with $N = 1, n = 1$.  According to Ref. [17], the numerical mass for this knot is $m \simeq 504$ e.u. $\simeq 3$ GeV.  Then the corresponding Compton wavelength is $\lambda_{c} = 4 \times 10^{- 14}$ cm.  For a rough estimate of the scattering cross-section, we follow Ref. [3] and write $\sigma = 4 \pi a^{2}$ where the scattering length $a \simeq 100 \lambda_{c}$.Then $\sigma \simeq 2 \times 10^{-22}$ cm$^{2}$, and $\sigma/m \simeq 6.6 \times 10^{-23}$ cm$^{2}$/GeV, a value that is within the range, $\sigma/m \simeq 8 \times 10^{-(25-22)}$ cm$^{2}$/GeV prescribed by Ref. [3].

Next, we consider the electroweak knot for which the energy unit is much larger than that for the nucleon scale (e.u.$ \simeq 5.9$ MeV).  Using the relations $\eta^{2} \kappa^{2} = 2/g^{2}$ and $M_{W} \simeq \eta g/2$ where $g = 3 \hbar c/2$ is the coupling constant, and $M_{W} \simeq 80$ GeV is the mass of the $W$ boson, we obtain $\eta^{2} \kappa^{2} \simeq 5.5 \hbar c$ and $\kappa \simeq 1.74 \times 10^{-16} $cm.  Thus, the electroweak energy unit is (e.u.)$_{ew} \simeq \eta^{2} \kappa/4 \simeq 1 55$ GeV.  Consequently, the mass of the electroweak knot, with $N= 1, n=1 [17]$, is $m \simeq 504 $(e.u.)$_{ew} \simeq 7.8 \times 10^{4}$ GeV.  Taking $\sigma = 4 \pi a^{2}$ where $a \simeq 10^{2} \kappa$, we obtain $\sigma/ m \simeq 5 \times 10^{-32}$ cm$^{2}$/ GeV, a value that is about $10^{-7}$ times the lower limit of the prescribed range [3].  Thus, owing to its large mass, the $N=1, n=1$ electroweak knot cannot be a candidate for the dark matter particle in the scenario of Ref. [3].

The circular loops studied in Secs. 7 and 8 are classically stable implying that the topological invariant $N$ is conserved.  However, this is not strictly true quantum mechanically, since there is a possibility of $N \rightarrow N-1$ transitions due to quantum tunneling.  This process involves migration across the loop of the flux line that is associated with the current along the loop.  We now consider the circular $ N= 1, n =1$ loop decaying into the $N = 0, n = 1$ loop.  To estimate the decay rate, we use the WKB formula $1 / \tau \simeq \omega_{0}$ exp $(-S_{E}) $ where $\omega_{0}$ is the rate of hitting the barrier and $S_{E} \simeq (2\kappa / \hbar) \sqrt{2M \Delta \epsilon}$. Here $M$ is the mass of the migrating line and $\Delta \epsilon \simeq \epsilon_{tot} \kappa$ is the energy of a loop segment of width $\kappa$.  For $N = 1, n =1$, we obtain $\omega_{0} \simeq c/(\sqrt {\pi} \kappa)$, $M \simeq 17 \eta^{2} \kappa/c^{2}$ and $\Delta \epsilon \simeq 20 \eta^{2} \kappa$, implying $S_{E} \simeq 52 \eta^{2} \kappa^{2}/(\hbar c)$.

For the nucleon scale loop, we have $\eta ^{2} \kappa^{2} \simeq 7.1 \times 10^{-2} \hbar c$, yielding $S_{E} \simeq 3.7$. With $\kappa \simeq 4 \times 10^{-14} $cm, the decay time is $\tau \simeq 2 \times 10^{-24}$ exp $(S_{E})$ s $ \simeq 8 \times 10 ^{-23}$s which is extremely short in comparison with the age of the universe $10^{17} $s.  This excludes the $N = 1, n=1$ knot from being the dark matter candidate within the scenario of Ref. [3]. 

This prompts us to examine the stability of the $N=0$ loops.  Since the analysis of the $N=0, n=1 $ case is complicated by the presence of logarithmically divergent integrals, we confine ourselves to the $N= 0, n=2 $ case.

First, we address the transverse stability condition for this case.  Letting $m=N=0$ in Eq. (72), and evaluating the integrals with the Ansatz (41), we obtain $\lambda^{2}_{b} \simeq 2 R \kappa$.  Next, we consider the loop energy $\epsilon_{loop} \simeq 2 \pi R \epsilon_{tot}$, where $\epsilon_{tot}$ is given by the expression (75) with $m=0$ and $\lambda^{2}_{b} \simeq 2 R \kappa$.  In this way, we obtain 

\begin{equation}\label{91}
\epsilon_{loop} = 8 \pi^{2}\eta^{2} \Big ( R + \kappa/2 + \kappa^{2}/(3R) \Big )
\end{equation}

 There is a remarkable similarity of this result to the expression (61).  We see that the third  term in the parenthesis of (91) now plays the role of the centrifugal energy which stabilizes the loop against a radial collapse.  The expression (91) has a minimum at the preferred radius given by $R_{e} \simeq \kappa / \sqrt{3}$.  Using this result in Eq. (91), we obtain $\epsilon_{loop} \simeq 130 \eta^{2} \kappa$.  For the nucleon scale loop, this implies a mass $\simeq 3$ GeV.  The $N = 0, n = 1$ loop is expected to have a lower mass owing to the decreased kinetic energy of the currents (see Eq. (19)).  With a mass that is smaller than 3GeV, the ratio $\sigma/m$ is expected to be within the range required by the dark matter scenario of Ref. [3].  

We should now inquire about the gravitational mass of the $N= 0, n=2$ loop.  Letting $N= 0$ in Eq. (90), we have $\bar {m} \simeq \pi^{2} \eta ^{2} \lambda^{2}_{b}/R ^{2}_{e}$.  Using $\lambda^{2}_{b} \simeq 2 R_{e} \kappa$ and $R^{2}_{e} \simeq \kappa^{2}/3$, we obtain from this result $\bar{m}_{a} \simeq 34 \eta^{2}$.  Then the gravitational mass of the $N=0, n=2$ loop becomes $\bar{M}_{a} = 2 \pi R_{e} \bar{m}_{a} \simeq 124 \eta^{2} \kappa$.  Hence, for the nucleon scale, $\eta^{2} \kappa \simeq 23.6$ MeV, we have $\bar {M}_{a} \simeq 2.9$ GeV.  Thus the gravitational mass of the $N=0, n=2$ loop is practically equal to the entire mass of the loop $\epsilon_{loop} \simeq 3.1$ GeV.  Consequently, these loops could play a role of the dark matter particles in the scenario of Ref. [3].

Although, the electroweak knots are too heavy yielding a ratio $\sigma/m$ that is seven orders magnitude below the range given in Ref. [3], they may still be of cosmological relevance.  We recall that in the electroweak scale, the parameters $\eta^{2}\kappa^{2} \simeq 5.5 \hbar c$ and $\kappa \simeq 1.7 \times 10^{-16} $ cm so that the action $S_{E} \simeq 286$.  Thus, the decay time of the $N=1, n=1$ electroweak knot is $\tau \sim 10 ^{113} $ s which is much longer than the age of the universe.  This is evidently due to the high energy cost for the intersection of the flux lines involved in the $N \rightarrow N-1$ transition.

For the same reason, the long electroweak twisted Baby Skyrmion strings are prevented from intercommuting.  This could lead to the formation of a frustrated string network. In an interesting paper, Bucher and Spergel [20] proposed that a solid dark matter component could arise from frustrated networks of non-Abelian cosmic strings.  Owing to the finite shear modulus, this solid component can have a negative pressure without instabilities that are expected for a perfect fluid.  Perhaps, the non-Abelian strings studied in the present paper could play a role in the theory of Ref. [20].


\begin{thebibliography}{17}


\bibitem{1}
L.~D. Faddeev,
\emph {Quantization of Solitons},
(Princeton preprint IAS-75-QS70).

\bibitem{2}
L.~D. Faddeev, A.~J. Niemi,
\emph{}
J. Phys. Lett. \textbf{82}, 1624 (1999).

\bibitem{3}
D.~N. Spergel and P.~J. Steinhardt,
\emph{}
Phys. Rev. Lett. \textbf{84}, 3760 (2000).

\bibitem{4}
A. Vilenkin and E.~P.~S. Shellard,
\emph {Cosmic Strings and Other Topological Defects},
(Cambridge Monographs on Mathematical Physics, Cambridge, 2004).

\bibitem{5}
E. Witten,  
\emph{}
Nucl. Physics B \textbf{249}, 557 (1985).

\bibitem{6}
R.~L. Davis and E.~P.~S. Shellard, 
\emph{}
Nucl. Phys. B \textbf{323}, 209 (1989); C.~J.~A.~P. Martins and E.~P.S. Shellard, \emph {}Phys. Rev. D \textbf{57}, 7155 (1998). 

\bibitem{7}
E. \v{S}im\'{a}nek,
\emph{}
Phys. Rev. D \textbf{78}, 045014 (2008).

\bibitem{8}
R.~A. Leese,
\emph{}
Nucl.Phys. B \textbf{366}, 283 (1991).

\bibitem{9}
Y. Verbin and A.~L. Larsen,
\emph{}
Phys. Rev. D \textbf{70}, 085004 (2004).

\bibitem{10}
A. Vilenkin, 
\emph{}
Phys. Rev. D \textbf{23}, 852 (1981).

\bibitem{11}
B.~M.~A.~G. Piette, B.~J. Schroers, and W.~J. Zakrzewski, 
\emph{}
Nucl. Phys. B \textbf{439}, 205 (1995).

\bibitem{12}
Y.~M. Cho, Hyojoon Khim, and Pengming Zhang,
\emph{}
Phys. Rev. A \textbf{72}, 063603 (2005).

\bibitem{13}
N. Manton and P. Sutcliffe,
\emph {Topological Solitons},
(Cambridge Monographs on Mathematical Physics, Cambridge, 2004).

\bibitem{14}
A. Einstein,
\emph {The Meaning of Relativity},
(Princeton University Press, Princeton, Newy Jersey, 1972).

\bibitem{15}
E. Copeland, M. Hindmarsh, and N. Turok,
\emph{}
Phys. Rev. Lett. \textbf{58}, 1910 (1987).

\bibitem{16}
L~D. Landau and E.~M. Lifshitz,
\emph {Theory of Elasticity},
(Pergamon Press, Oxford, England, 1975).

\bibitem{17}
R.~A. Battye and P.~M. Sutcliffe,
\emph{}
Phys. Rev. Lett. \textbf{81}, 4798 (1998).

\bibitem{18}
G.~S. Adkins, C.~R. Nappi, and E Witten,
\emph{}
Nucl. Phys. B \textbf{228}, 552 (1983).

\bibitem{19}
J. Gladikowski and M. Hellmund,
\emph{}
Phys. Rev D \textbf{56}, 5194 (1997).

\bibitem{20}
M. Bucher and D. Spergel,
\emph{}
Phys. Rev D \textbf{60}, 043505 (1999).





\end{thebibliography}
\end{document}